\documentclass[twocolumn]{aastex631}
\usepackage{gensymb}
\usepackage{xcolor}
\usepackage{subfigure}
\newcommand{\anu}[1]{\textcolor{black}{{#1}}}
\newcommand{\bbk}[1]{\textcolor{black}{{#1}}}

\newcommand{\Fig}[1]{Figure~\ref{#1}}

\newcommand{\Sec}[1]{Section~\ref{#1}}

\newcommand{\Tab}[1]{Table~\ref{#1}}

\newcommand{\App}[1]{Appendix~\ref{#1}}
\newcommand{\Rsun}{${\rm R_\odot}$}
\def\ndash{\,--\,}
\def\bmax{$B_{\rm max}$}
\def\bmean{$B_{\rm mean}$}

\shorttitle{Automatic Algorithm for BMR Tracking}
\shortauthors{Sreedevi et al.}

\graphicspath{{./}{figures/}}
\begin{document}

\title{AutoTAB: Automatic Tracking Algorithm for Bipolar Magnetic Regions}

%DTAB: Detection and Tracking Algorithm for Bipolar Magnetic Regions\\
%AutoTAB is good but does not capture the full form of BMR. What about this:
%AATBMR: Algorithm for Automatic Tracking  of Bipolar Magnetic Regions

\author[0000-0001-7036-2902]{Anu Sreedevi}
\affiliation{Department of Physics, Indian Institute of Technology (Banaras Hindu University), Varanasi 221005, India}
\email{anubsreedevi.rs.phy20@itbhu.ac.in}

\author[0000-0003-3191-4625]{Bibhuti Kumar Jha}
\affiliation{Southwest Research Institute, Boulder, CO 80302, USA}
\affiliation{Aryabhatta Research Institute of Observational Sciences, Nainital 263002, Uttarakhand, India}

\author[0000-0002-8883-3562]{Bidya Binay Karak}
\affiliation{Department of Physics, Indian Institute of Technology (Banaras Hindu University), Varanasi 221005, India}
\email{karak.phy@iitbhu.ac.in}

\author[0000-0003-4653-6823]{Dipankar Banerjee}
\affiliation{Aryabhatta Research Institute of Observational Sciences, Nainital 263002, Uttarakhand, India}
\affiliation{Indian Institute of Astrophysics, Koramangala, Bangalore 560034, India}
\affiliation{Center of Excellence in Space Sciences India, IISER Kolkata, Mohanpur 741246, West Bengal, India}

\begin{abstract}
Bipolar Magnetic Regions (BMRs) provide crucial information about solar magnetism. They exhibit varying morphology and magnetic properties throughout their lifetime, and studying these properties can provide valuable insights into the workings of the solar dynamo. The majority of previous studies have counted every detected BMR as a new one and have not been able to study the full life history of each BMRs. To address this issue, we have developed an Automatic Tracking Algorithm (AutoTAB) for BMRs,  that tracks the BMRs for their entire lifetime or throughout their disk passage. AutoTAB uses the binary maps of detected BMRs to automatically track the regions.
This is done by differentially rotating the binary maps of the detected regions and checking for overlaps between them. 
In this first article of this project, we provide a detailed description of the working of the algorithm and evaluate its strengths and weaknesses by comparing it with existing algorithms. AutoTAB excels in tracking even for the  small BMRs (with flux $\sim 10^{20}$ Mx) and it has successfully tracked 9152 BMRs over the last two solar cycles (1996--2020), providing a comprehensive dataset that depicts the evolution of various properties for each BMR. The tracked BMRs exhibit the well-known latitudinal-time distribution and 11 year cyclic variation except for small BMRs which appear at all phases of the solar cycle and show weak latitudinal dependency.
Finally, we discuss the possibility of adapting our algorithm to other datasets and expanding the technique to track other solar features in the future.
\end{abstract}


\section{Introduction} 
\label{sec:intro}
Bipolar Magnetic Regions (BMRs) remain one of the most predominant 
signatures of solar magnetism as observed on the surface of the Sun.
The number of such regions at a given time represents the solar activity 
that waxes and wanes cyclically over a period of 11~years \citep[]{Schwabe1844, Hathway2015}. These BMRs are generally the source regions 
of solar eruptive events \citep{Schrijver2009}, and hence they are crucial for the understanding 
of space weather conditions. BMRs are observed to be tilted  by an angle with respect to the equator, which is found to increase statistically with latitude known as Joy's law \citep{Hale1919}. It is observed that the decay 
and dispersal of tilted BMRs produce the poloidal field in the Sun through  Babcock\--Leighton mechanism \citep{Babcock1961, Leighton1964, Mord22}. 

Sunspot groups \anu{in white light observations, mimic} the locations of large BMRs, and have been traditionally used for the study of BMR properties \citep[e.g.,][]{Howard1991, Howard96, Sivaraman2007, Dasi-Espuig2010} because of the unavailability of magnetogram data. Since the early 1970s, regular full-disk measurements of the Sun's magnetic field have been started, and now there exist vast archived data of solar magnetograms from the different ground-based (Synoptic Optical Long-term Investigations of the Sun \citep[SOLIS;][]{Keller2003}, Global Oscillation Network Group \citep[GONG;][]{Harvey1996}) and space-based (Michelson Doppler Image \citep[MDI;][]{Scherrer1995}  and Helioseismic and Magnetic Imager \citep[HMI;][]{Scherrer12}) observatories. To exploit such a large volume of data for the understanding of BMRs and solar magnetism, various 
automatic methods has been developed \citep{Stenflo2012, Tlatov2014, Jha2020}. 
 In all these studies, each detection of BMRs has been treated as a new one \citep{Stenflo2012, Jha2020}. However, this approach may influence the analysis as bigger BMRs have a longer lifetime and thereby will have higher weightage in the analysis \citep{Lidia2015}. Furthermore, BMR properties, including its morphology, magnetic field strength, and tilt angle, evolve significantly over its lifetime \citep{Ugarte-Urra15, McClintock2016, Getling19, Schunker2019, Schunker2020}. Hence, to overcome these limitations,
it is essential to track the BMR to get insight into the physics of the formation and evolution of BMR. 
Moreover, the automatic detection and tracking of magnetic regions also become essential for monitoring solar activity and events \citep{McAteer2005, Labonte2007}.
Therefore, an enormous effort has been made to develop automatic algorithms to track the BMRs \citep{Higgins2011, Jaramillo2016}. 

Solar Monitor Active Region Tracking \citep[SMART;][]{Higgins2011} was developed for the automatic detection and tracking of magnetic active regions in real-time for solar eruptive event prediction. Although SMART does a decent job in identifying and extracting various magnetic features, it tends to miss some quiet Sun magnetic regions. The Bipolar Active Region Detection \citep[BARD;][]{Jaramillo2016} is an another algorithm developed for the detection
and tracking of BMRs. It uses similar techniques to detect BMRs as SMART but uses the dual-maximum ﬂux-weighted overlap method for feature association. The BARD uses human supervision to correct any errors in detection and tracking. Furthermore, Space-Weather MDI Active Region Patches \citep[SMARP;][]{Bobra2021} and  Space-Weather HMI Active Region Patches \citep[SHARP;][]{Bobra2014} are the data sets derived from MDI and HMI magnetograms, respectively which provide the tracked maps of active regions identified in the magnetic image of the Sun. We note that in SMARP and SHARP, active regions are not necessarily the BMRs, where a decent flux balance condition holds.

To the best of our knowledge, there is no existing catalog that can provide the properties of the individual BMRs throughout their lifetime or at least
the course of their lifetimes. Therefore, we overcome the limitation of existing BMR tracking algorithms and develop a completely automatic method to track the BMR, which can be implemented in all sets of magnetogram data. In \Sec{sec:data}, we present a detailed description of our tracking algorithm and its comparison with the existing one. In \Sec{sec:res}, we present some representative results based on our tracking algorithm, and finally, in \Sec{sec:conclusion}, we conclude with our insight on this new algorithm.

\section{Data and Method} \label{sec:data}

In our study, we use the Line-of-Sight (LOS)
\anu{magnetic field observation of the Sun usually referred as magnetogram,} for the period of September 1996 to December 2019 (Cycles~23 and 24) from 
%Michelson Doppler Imager \citep[MDI: 1996--2011;][]{Scherrer1995}
MDI \citep[1996--2011;][]{Scherrer1995}
 and %Helioseismic and Magnetic Imager \citep[HMI: 2010--present;][]{Scherrer12}
HMI 
\citep[2010--present;][]{Scherrer12}
onboard the Solar and Heliospheric Observatory(SOHO) and Solar Dynamic Observatory (SDO), respectively. Here, we utilize all the magnetogram \anu{data} from MDI, which comes with a cadence of 96~minutes in 1024$\times$1024 resolution with a %pixel size of 2\arcsec $\times$ 2\arcsec ~(
spatial resolution of  4\arcsec $\times$ 4\arcsec. Although HMI provides magnetogram with the cadence of 45~sec in 4096$\times$4096 resolution with %the pixel size of 0.5\arcsec $\times$ 0.5\arcsec ~(
spatial resolution of  1\arcsec $\times$ 1\arcsec,  for the ease of computation, we only use one image every 96~minutes as available for MDI, from HMI
series ``hmi.M\_720s.''
%for the period of 2010\,--\,2022.

The data sets used for the mentioned period contain 1,26,381 fits images (MDI:~56,384 HMI:~69,997). In addition to quality keyword, number of faulty pixels (identified as ``Not a Number'') on the solar disk \anu{region} is also checked for every magnetogram. If the number of faulty pixels are found to be greater than 100, the magnetogram is considered defective and not included in the analysis.

%Data in the magnetograms essentially represents the LOS component of the surface magnetic field integrated over spatial resolution kernel. 
The data in the magnetograms corresponds to the 
%line-of-sight (LOS) 
LOS
%LOS is already defined
component of the surface magnetic \anu{flux density}, %field, which is
\anu{calculated} across a spatial resolution \anu{window}.%kernel.
%\anu{at the pixel level}. Hence, 
 \anu{ These LOS magnetograms suffer} from the projection effects, which is inversely proportional to the cosine of heliocentric angular distance $(\mu)$\footnote{Angular distance of the pixel in consideration from the disk center.}. We correct the LOS component for this effect by assuming the magnetic field is normal to the solar surface and hence, \anu{dividing surface magnetic field density at each pixels by $\cos{\mu}$}. \anu{Furthermore, as we approach to limb the projection effects become severe and hence,} initially we restrict ourselves to less than 0.9~\Rsun. 
%adding 'less than here'
%\textcolor{red}{Furthermore, we also noticed that as we go close to the limb, multiple closely lying BMRs are identified as one by the detection \citep{Jha2020}.}

%As this may significantly affect the tracking, we later restrict our analysis only between $\pm45\degree$ of longitudes.

\begin{figure*}[!htbp]  
\includegraphics[width=2\columnwidth]{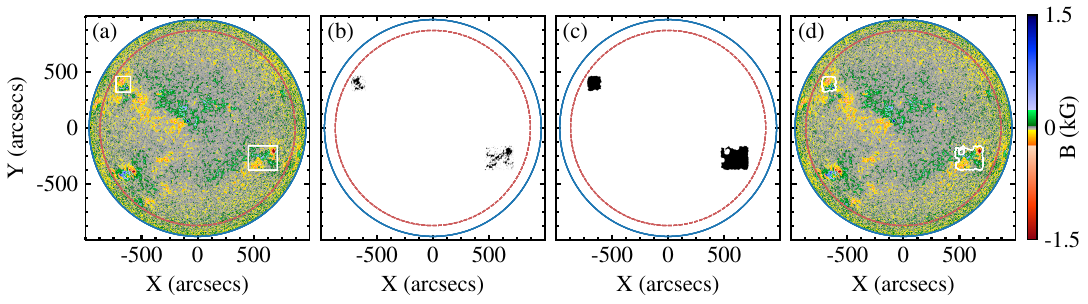}
\caption{
(a) Magnetogram \anu{from MDI} for 15-11-2000 20:48 with the detected BMRs represented by white rectangular boxes. (b) %Same as
\anu{Binary mask of detected BMRs from (a) but after applying the threshold of 100~G to the detected regions. (c) Same as (b) after the pre-processing (described in \Sec{preprocess}). (d) Identified regions after pre-processing depicted on the magnetogram.} %Same as (a) after preprocessing.} of BMR masks. 
The solid blue and red dashed circles represent 1.0~\Rsun\ and 0.9~\Rsun, respectively.}
\label{fig:fig1}
\end{figure*}

\subsection{BMR Detection Algorithm}
\label{section:det_alg}
Algorithm developed to detect BMRs from the LOS magnetograms follows the prescription given in \citet{Stenflo2012}, which was also used in \citet{Jha2020} with slight modifications. A brief description of the detection algorithm is as follows. %\textcolor{red}{In the first step, the LOS magnetograms are corrected for projection effects to obtain the radial component of the solar magnetic field.} %magnetic field component normal to the solar surface 
\anu{In the first step, a adaptive threshold, based on the average of absolute magnetic field inside the $0.9$~\Rsun\ is applied on the projection corrected magnetogram, to identify the regions of strong magnetic fields. This step is followed by applying a threshold of 200~G to identified regions and imposing a flux balance condition defined by  $(F_{+} + F_{-})/(F_{+} - F_{-}) < 0.4$ on them to isolate the BMRs. Here, $F_{+}$ and $F_{-}$ represent the sum of all the positive and negative radial magnetic field values, respectively, that exceed 100~G for the identified regions  \citep[see Appendix of][for detail description of the detection method]{Stenflo2012}. } 
%After that, an appropriate threshold is used for every magnetogram based on its average magnetic field value to isolate the strong magnetic field regions.  The identified BMRs are shown in \autoref{fig:fig1}a. 
\anu{\citet{Stenflo2012} has only used this algorithm for MDI data set and it has been extended for HMI in \citet{Jha2020}. \citet{Jha2020} has noted that the same method could be used for HMI data by multiplying the HMI magnetogram by the factor of 1.4, which comes from the comparison of MDI and HMI magnetogram in the overlapping period \citep{Liu2012}. 
% Although, \citet{Liu2012} noted that the scaling factor varies over the disk but that variat  the factor of 1.4 neethresholds for the detection algorithm has been optimized for MDI magnetograms. 
To keep the consistency in the dataset,}
%In the case of 
for HMI magnetograms, we re-binned the magnetogram from $4096\times4096$~pixels to $1024\times1024$~pixels to further reduce the computational time. By comparing the detected BMRs from the original and the binned data, we find that \anu{binning tends to miss the detection of small BMRs with the flux of $\sim10^{20}$~Mx, but the detection of other BMRs remains unaffected.} %this binning does not affect the detection of BMRs. 
One such example of identified regions from MDI is shown in \autoref{fig:fig1}a. The BMR from both the data sets are stored in the form of a binary mask which acts as the starting point of our tracking algorithm.

\subsection{Pre-processing of BMR Masks}
\label{preprocess}
%In \citet{Jha2020}, the identified masks of BMRs are stored as the rectangular regions around the BMRs 
In \citet{Jha2020}, the identified BMRs are stored as rectangular region by binary masks (see \Fig{fig:fig1}a), which include additional pixels which are not part of BMRs and may affect the tracking, particularly in the case of closely spaced BMRs during high activity period. Hence, to get rid of those extra pixels before we start the tracking, we go through the following pre-processing steps.

To get the exact morphology of the BMR in the rectangular regions of interest (ROI), we picked all the pixels with \anu{absolute magnetic field values greater than 100~G}. This threshold not only leads to a few fragmented pixels in the ROI but also separates the poles of the BMR, as they do not always touch each other's boundaries (see \Fig{fig:fig1}b). To eliminate the fragmented pixels, and connect isolated parts, we use a technique called morphological closing. It works by filling in the gaps or holes in the image using a kernel with a certain shape and size\footnote{We use morph\_close.pro function available in IDL for this.
}. In our case, we start with a circular kernel that has a initial radius of 6 pixels, followed by an area threshold of 50 pixels for filled regions. The radius of the kernel is systematically increased up to 9 pixels at the step of one pixel until the number of connected regions matches with the number of identified BMRs (i.e., the number of rectangular regions, see  \Fig{fig:fig1}a) in that magnetogram. \anu{Finally, the flux balance condition, as described previously, \citep{Stenflo2012, Jha2020} is verified for each region to ensure that they are bipolar. In \Fig{fig:fig1}c, we show the BMR regions after morphological closing operation.} In \autoref{fig:fig1}d, 
white contours represent the BMRs after the pre-processing steps. The information of detected BMRs after pre-processing is also stored as binary masks so that it can be used for tracking.

\begin{figure*}[!htbp]
\centering
\includegraphics[width=\textwidth]{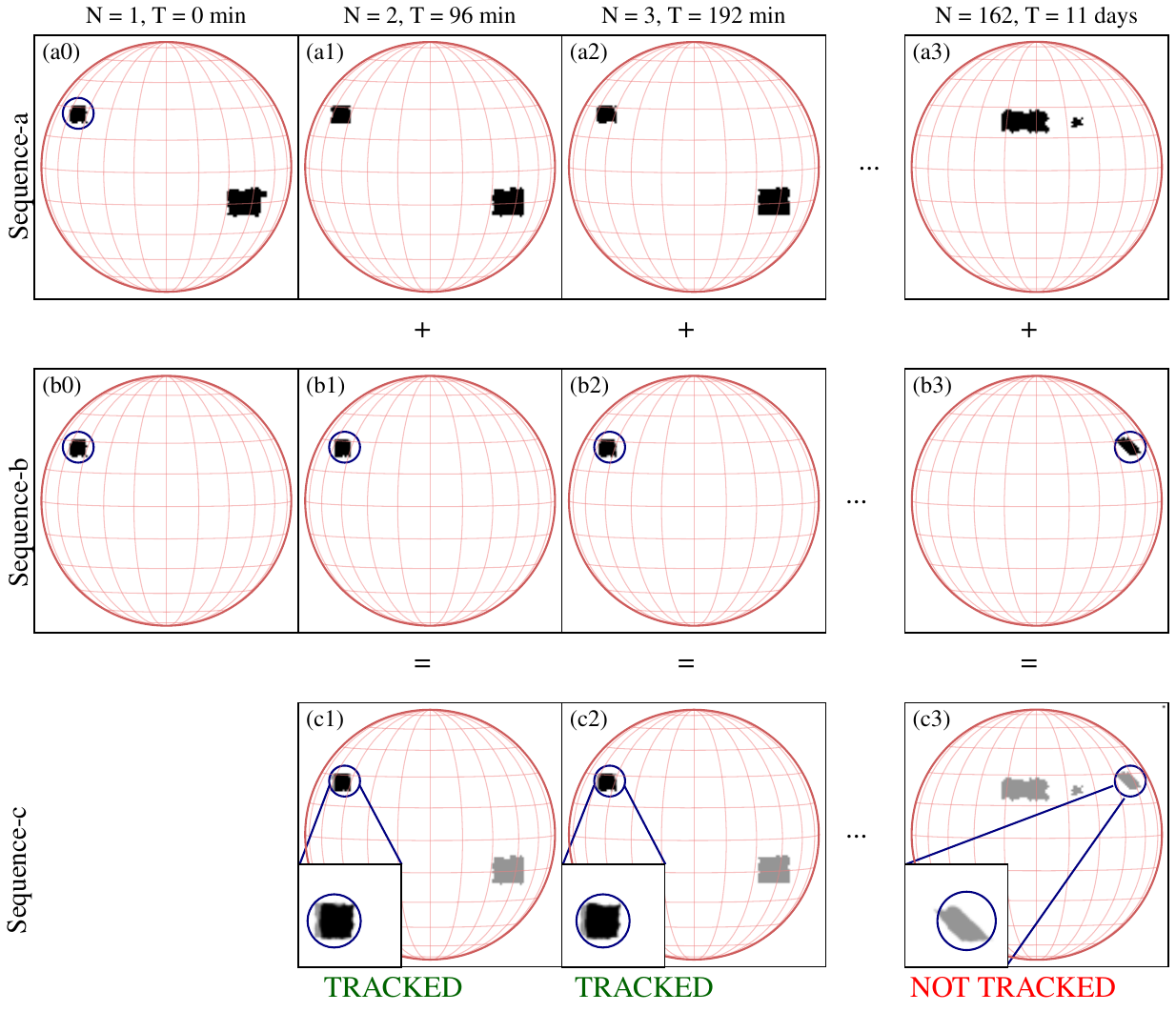}
\caption{Representative example of the BMR Tracking Algorithm. The BMR intended to be tracked is marked by a circle. (a0), (a1), (a2) and (a3) represent the selected binary maps in Sequence-a.
% the observations in the range of $T_{\rm max}$ of the BMR (in this particular case,  $T_{\rm max} = 11$~days with a cadence of 96 minutes.
(b0), (b1), (b2) and (b3) represent Sequence-b, the isolated and  differentially rotated binary masks of the BMR, corresponding to the time of observation in Sequence-a. (c1), (c2) and (c3) represent Sequence-c, obtained by adding Sequence-a and Sequence-b. The zoomed-in view of overlapping region is shown in the inset.}
\label{fig:fig2}
\end{figure*}

\subsection{BMR Tracking Algorithm} 
\label{sec:track_alg}

Now we come to the core of the work, and that is to develop an automatic algorithm to track all of the identified BMRs. Our tracking algorithm employs binary masks obtained from the last step of pre-processing to follow the interested regions over their lifetime/disk passage. The concept of the BMR tracking algorithm comes from the sunspot tracking algorithm developed in \citet{Jha2021}, which is modified considerably to track the BMRs in the magnetograms. Now, we discuss the steps of our tracking algorithm as follows.

\begin{enumerate}
   \item At the first step, a binary mask is selected from the archive (\autoref{fig:fig2}a1). A BMR \anu{is selected from this mask and checked for tracking history stored in step 7. If the selected BMR} is not tracked already (see the marked circle in \autoref{fig:fig2}a1), it is isolated in a separate binary mask (\autoref{fig:fig2}b1), and a unique BMR-ID (say 10001) is assigned to it. 
   % \anu{The information about the first detection of the BMR along with the assigned BMR-ID, is stored in a previously empty text file.} 

    \item Now, we calculate the maximum tracking period ($T_{\rm max}$) i.e., the time it takes to reach the west limb,  based on the heliographic latitude $\theta_{\rm  BMR}$ and longitude $\phi_{\rm BMR}$ of the BMR as,
    \begin{equation}
        T_{\rm max} = \frac{90\degree-\phi_{\rm BMR}}{\Omega(\theta_{\rm BMR})}.
    \end{equation}
    Here, $\Omega (\theta_{\rm BMR})$ is the photospheric differential rotation rate \citep{Howard1990} for given $\theta_{\rm  BMR}$.

    \item In the next step, we take a sequence of \anu{binary masks} %observations (binary masks; 
    (say Sequence-a, as  shown in \autoref{fig:fig2}a0\,--\,a3) falling in the range of $T_{\rm max}$.
    
    \item The binary mask of isolated BMR (\Fig{fig:fig2}b1) is differentially rotated using the drot\_map.pro routine in IDL\footnote{It uses differential rotation profile from \citet{Howard1990}.}, to the time of observation of Sequence-a. For example, b0 is differentially rotated to the time of observation of a1 to obtain b1; similarly, b0 is differentially rotated to the time of a2 to obtain b2, and so on.
    %(b2 = b1 $\rightarrow$ a2), similarly, b3 = b1$\rightarrow$a3 and so on. 
    This is represented by Sequence-b and is shown in \autoref{fig:fig2}b0\,--\,b3. 
    
    \item In the following step, we add the Sequence-a  (\autoref{fig:fig2}a0\,--\,a3) with the corresponding differentially rotated binary masks, Sequence-b (\autoref{fig:fig2}b0\,--\,b3),  e.g., c1=a1+b1, c2=a2+b2 and so on. This is represented by Sequence-c and is shown in \Fig{fig:fig2}c1\,--\,c2. 
    
    \item Now, we go through all the images in Sequence-c (\autoref{fig:fig2}c1\,--\,c3) and check for the overlapping pixels \anu{(pixels with value 2)}. If the overlap is more than 150~pixels ($600$~arcsec$^2$), the BMR is marked as tracked (\Fig{fig:fig2}c1 and \Fig{fig:fig2}c2), otherwise not tracked (\Fig{fig:fig2}c3). Furthermore, if the overlapping criteria are not met in the consecutive 30 binary masks
    (i.e., not enough overlap was detected for the next 48 hours),
    we increase the overlapping criteria to 250 pixels (1000~ arcsec$^2$) for the next observation to ensure that the BMR tracked is the intended one and not a new one.

    \item The same BMR-ID i.e., 10001, is assigned to all the tracked BMRs in Sequence-c. At this step, along with BMR-ID, we also store a few other parameters of the tracked BMR  \anu{such as \bmax, \bmean, total unsigned flux, area, and heliographic coordinates.} \anu{The time and unique region index, which is assigned to each connected regions (pixels with value 2) in Sequence-c along with the BMR-ID are appended in the tracking history file, which will be used at step 1 to check for tracking history.} 

    \item We repeat all the above steps for \anu{all the other detected BMRs within the chosen binary map from Step-1 (\autoref{fig:fig2}a0).} 
    %each BMR present in the selected observation period (\autoref{fig:fig2}a1).

    \item Once all the BMRs in the first selected observation are tracked, we go for the next observation and track only the new BMRs which are not tracked already. \anu{We do this by comparing BMR-ID with the existing information in the tracked history file updated} in the previous step.
\end{enumerate}

\section{Results}
\label{sec:res}

\subsection{Representative Tracking Results}

\begin{figure*}[!htbp]
\gridline{\fig{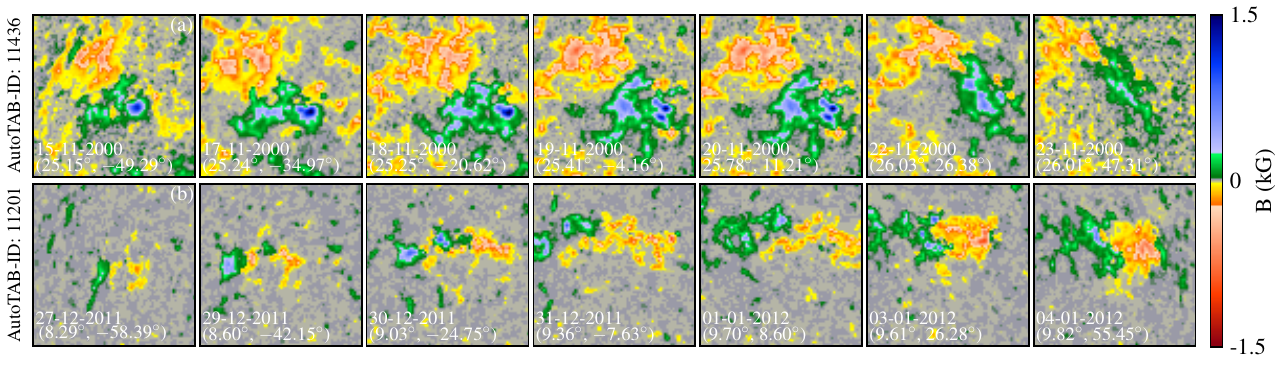}{\textwidth}{}}
\gridline{\fig{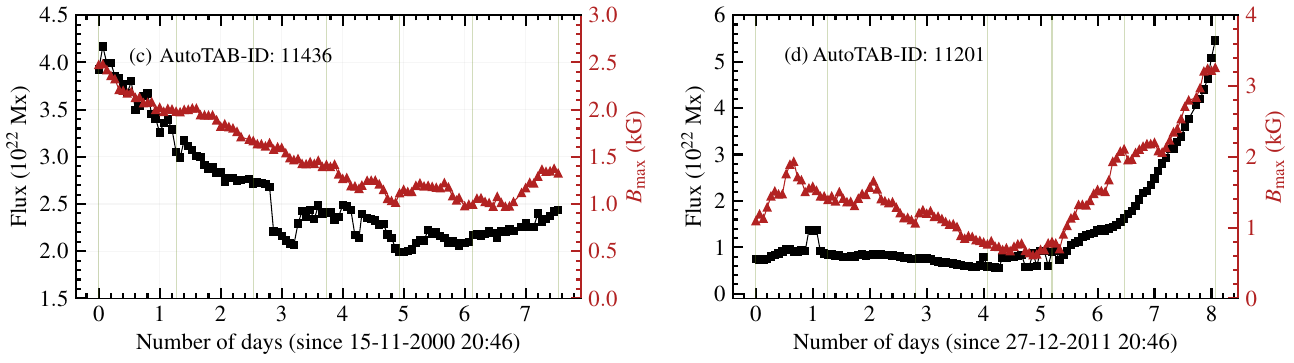}{\textwidth}{}}
     % \centering
     % \begin{subfigure}{1.0\textwidth}
     %     \centering
     %     \includegraphics[width=\textwidth]{fig_31.pdf}
     %   \end{subfigure}
     % \hfill
     % \begin{subfigure}{1.0\textwidth}
     %     \centering
     %     \includegraphics[width=\textwidth]{fig_32}
     % \end{subfigure}
     \caption{
     Panels (a) and (b) respectively show the snapshots of the tracked BMRs with AR9232 (AutoTAB-ID 11436) and AR11390 (AutoTAB-ID 11201) corresponding to each day during the tracking. Panels (c) and (d) show the evolution of absolute total flux and \bmax\ for these BMRs. The vertical lines in (c) and (d) represent the times corresponding to the time of snapshots shown in (a) and (b), respectively.}
     \label{fig:fig3}
\end{figure*}

To demonstrate the result of our tracking algorithm, in \autoref{fig:fig3}, we show the evolution of NOAA region AR9232 (AutoTAB-ID 11436) and AR11390 (AutoTAB-ID 11201), which have been observed in MDI and HMI, respectively. AR9232 was first identified near the East limb ($25.2\degree$~N, $49.3\degree$~E) and has been tracked throughout its disk passage from 18-06-2000 03:11 to 24-06-2000 22:23\footnote{All the mentioned time in this article is in UTC.}. Although our algorithm tracks the AR9232 in \anu{next 113 timesteps with 96~minute cadence}, 
%113 observations (images), 
in \autoref{fig:fig3}a, we only show one snapshot every day for 7~days. In \autoref{fig:fig3}c, we also show the evolution of \bmax\ and absolute total flux for AR9232, and we note a systematic and continuous decrease (except small fluctuation \anu{due to fragmentation of  the negative polarity}) in both these quantities. This continuous decrease suggests that AutoTAB has been able to track the BMR properly, and it has been picked during its decaying phase, which is also inferred from  \autoref{fig:fig3}.
In contrast with our algorithm, the SMART Algorithm \citep{Higgins2011} tracks the same region
% for a shorter period, 
from 15-11-2000 03:15 to 23-11-2000 20:48.

In \autoref{fig:fig3}b, we show the evolution of AR11390 observed in HMI in a similar way as for AR9232. This BMR has been first identified near the East limb ($8.3\degree$~N, $58.2\degree$~E), and AutoTAB tracks it for 8~days (in 122 observations) till it reaches close to the West limb. The variation of total flux and \bmax\ (\autoref{fig:fig3}d) once again show the continuous and systematic variation over the tracking period. Furthermore, in \autoref{fig:fig3}d, we see that the \bmax\, as well as absolute total flux, show a small variation initially, and then they start increasing rapidly on day 5. Hence, we infer that this BMR has been tracked from its emergence to the time it crosses the West limb when AutoTAB loses track of it.

Apart from these two BMRs, in \App{sec:app}, we also show the evolution of another tracked BMR which has been observed in both MDI and HMI
%exist
(also see \Fig{fig:fig9}). The comparison of \bmax\ and total flux of this BMR obtained from MDI and HMI confirms that the AutoTAB is working in the same way for both the data sets. Please note that we have scaled up the \bmax\ measured in HMI by 1.4 to bring both MDI and HMI at same level \citep{Liu2012}.

By selecting the BMRs randomly from the tracked data, we observe that similar kinds of variations are consistently present in all the tracked BMRs with a few exceptions. 
Therefore, based on these findings, we say that AutoTAB is very efficient in tracking the BMRs during their appearance on the near side of the Sun. Using our state-of-the-art algorithm, we have tracked %8809
9232 BMRs in Cycle-23 and Cycle-24 for the years 1996\ndash2022. In the following section, we will discuss some of the statistical properties of these tracked BMRs.

\begin{figure*}[htbp!]
\centering
\includegraphics[width=\textwidth]{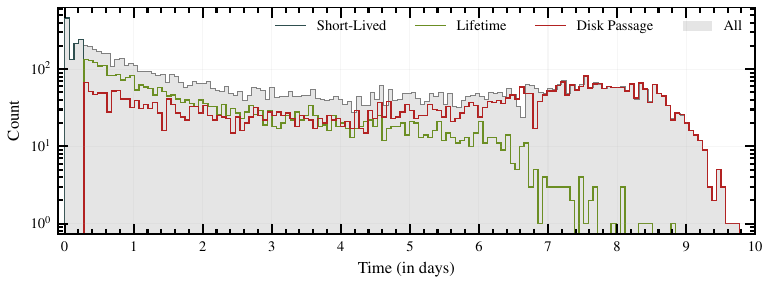}
\caption{Distribution of the lifetimes of the tracked BMRs of different classes shown by different colors.}
\label{fig:fig4}
\end{figure*}

\subsection{Statistical Properties of BMRs}
\label{sec:stat_prop}

In \autoref{fig:fig4}, we show the distribution of the lifetime/disk-passage time for all the tracked BMRs. From \autoref{fig:fig4}, we note that BMRs exhibit a broad range of lifetimes, varying from several hours to over a week. Although AutoTAB is not able to track a BMR that goes into the far side, it has been tracked only in its evolutionary stage. Other than that, we also encountered a few cases in which the BMR is living for a sufficiently long time, but they have been detected for less than 5\% of times, in the  number of expected frames\footnote{Number of expected frames = (Time of last detection $-$ Time of first detection) / cadence of data.} due to data gaps, corrupted data, or they have been missed during detection. 
Hence, after excluding such cases, based on how they were tracked i.e., either for the whole life or only in the evolutionary stage, we classified the BMRs into three classes, which are as follows. 

\begin{enumerate}
    \item \textbf{Short Lived (SL):} The BMRs, which emerge and decay in the near side of the Sun and have a lifetime of 8 hours or less, are classified as Short Lived (SL) BMRs. We segregated this class as they mainly comprise small BMRs, which may include a few large ephemeral regions with the typical magnetic flux of $\approx10^{20}$~Mx. The distribution of the SL class appears in the left most of \autoref{fig:fig4}. In \autoref{fig:fig5}a, we show the evolution of total flux and \bmax\ of a typical BMR (AutoTAB-ID 10076) from this class.  It is noted from  this figure that the BMR shown here emerges with relatively low flux and \bmax\ and decays within a few hours.

\begin{figure}[!htbp]
    \centering
    \includegraphics[width=\columnwidth]{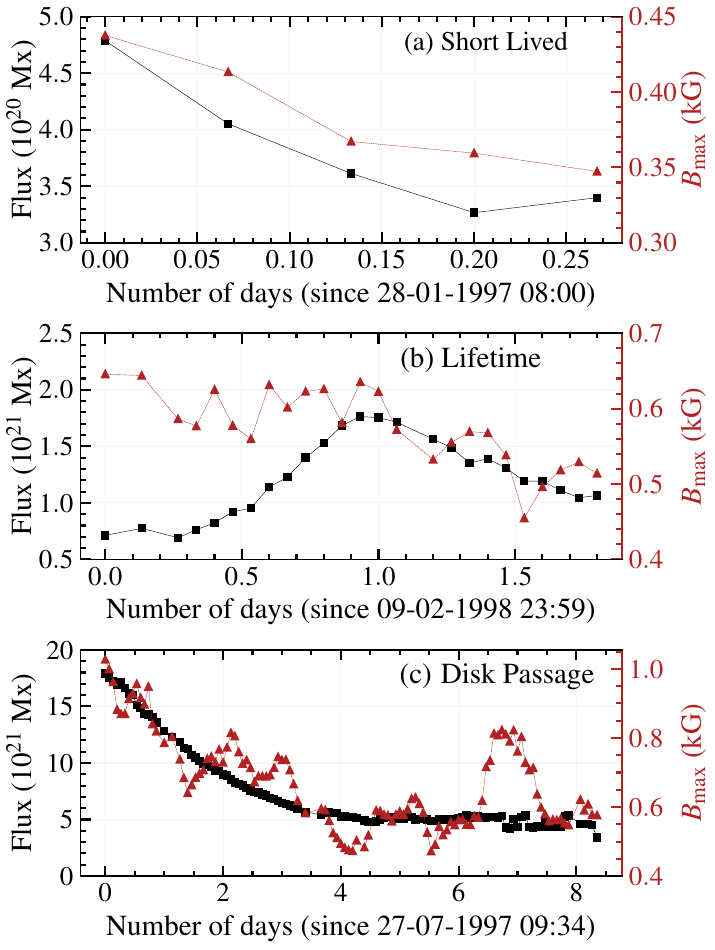}
    \caption{Evolution of absolute total flux and \bmax\ for a typical BMRs from (a) SL, (b) LT and (c) DP classes.  AutoTAB-IDs of the BMRs shown in (a), (b), and (c) are 10076, 10399, and 10213, respectively. \anu{NOAA number assigned only to the representative BMR of DP class is  AR08066.}}
    \label{fig:fig5}
\end{figure}

    \item \textbf{Lifetime (LT):} Rest of the tracked BMRs, which have a lifetime of more than 8~hours and have emerged and decayed in the near side of the Sun, i.e., they have been tracked throughout their lifetime, are classified as Lifetime (LT) BMRs. The distribution of the LT class spans around a week which is shown in  \autoref{fig:fig4}. An example of the evolution of such BMR (AutoTAB-ID 10213) is shown in \autoref{fig:fig5}b.

    \item \textbf{Disk Passage (DP):} This class of BMRs has not been tracked for their lifetime. Instead, they have only been tracked for a part of their life span on \anu{the visible solar disk,}
    %. This class includes all those BMRs, 
    and we classed them as Disk Passage (DP). This class includes $(i)$ BMRs that appear near the East limb ($\le45\degree$E) and disappear on the near side, $(ii)$ BMRs that appear on the near side of the Sun but cross the West limb (longitude $\ge45\degree$~W) and $(iii)$ BMRs which appears on the East limb ($\le45\degree$~E) and crosses West limb ($\ge 45\degree$W). Here we restrict ourselves between the longitudes of $\pm45\degree$, as the uncertainty in the magnetic field measurement increases towards the limb. Lifetime for this class is distributed all the way from a few hours to more than a week \autoref{fig:fig4}. The evolution of flux and \bmax\ for one such BMR (AutoTAB-ID 10399; \anu{NOAA AR08066}) is shown in \autoref{fig:fig5}c.

\end{enumerate}
The snapshots of the evolution of all the three BMRs in \autoref{fig:fig5} can be found in the data repository\footnote{https://github.com/sreedevi-anu/AutoTAB} for AutoTAB.

 The number of BMRs identified in each class is represented in the \Tab{tab:tab1} for the period of 1996\ndash2020. Along with their numbers, we also calculated the mean of area, total flux, \bmax\ and \bmean\ for all these classes, which are listed in \autoref{tab:tab1}. From this table, we note that all these quantities increases as we go from SL class to LT and DP classes. Hence, we find that the bigger BMRs have longer lifetimes, which is in agreement with the earlier findings from the sunspot \citep{Lidia2015}. 
 %\bkjq{ref}.  

 \begin{deluxetable*}{lcrcrc}%[!t]
\tablecaption{Some key parameters of different classes of tracked BMRs.
\label{tab:tab1}}
\tablehead{\colhead{Classification} & \colhead{Number of BMRs}  & \colhead{Area $\pm\Delta$Area}  & \colhead{Flux $\pm\Delta$Flux} & \colhead{\bmax $\pm\Delta$\bmax}  & \colhead{\bmean $\pm\Delta$\bmean}\\
\colhead{} &\colhead{} & \colhead{($\mu$Hem)} & \colhead{($10^{22}$Mx)} & \colhead{(G)} & \colhead{(G)}}
\startdata
Short Lived (SL)  & 1251  & 20.17 $\pm$ 0.71  & 0.26 $\pm$ 0.01  & 541.32 $\pm$ 5.38 & 197.46 $\pm$ 0.66\\ 
Lifetime (LT)     & 3191  & 88.65 $\pm$ 1.05  & 1.50 $\pm$ 0.02  & 949.20 $\pm$ 6.35 & 224.46 $\pm$ 0.81\\
Disk Passage (DP) & 4710  & 116.87 $\pm$ 0.17  & 2.05 $\pm$ 0.01  & 1436.83 $\pm$ 7.02 & 281.07 $\pm$ 0.71
\enddata
\end{deluxetable*}

%\begin{deluxetable*}{lcrcr}[h]
%\tablecaption{Distribution of BMRs into different categories  
%\label{tab:tab1}}
%\tablehead{\colhead{Classification} & \colhead{Number of BMRs}  & \colhead{Area + $\Delta$Area}  & %\colhead{Flux + $\Delta$Flux} & \colhead{\bmax + $\Delta$\bmax}\\
%\colhead{} &\colhead{}& \colhead{($\mu$Hem)} & \colhead{($10^{22}$Mx)} & \colhead{}}
%\startdata
%Short Lived (SL)  & 1150  & 21.43 $\pm$ 0.78  & 0.34 $\pm$ 0.01  & 555.42 $\pm$ 5.83 \\ 
%Lifetime (LT)     & 2974  & 94.16 $\pm$ 1.13  & 1.60 $\pm$ 0.02  & 990.33 $\pm$ 6.81 \\
%Disk Passage (DP) & 4607  & 118.96 $\pm$ 0.95  & 2.09 $\pm$ 0.02 & 1454.12 $\pm$ 6.95\\
%\enddata
%\end{deluxetable*}

After discussing the various classes of BMRs and their time evolutions, we shall now discuss
the collective behavior of the BMRs in the aforementioned three classes.

\subsubsection{Solar Cycle Variation}
\begin{figure}[htbp!]
\centering
\includegraphics[width=\columnwidth]{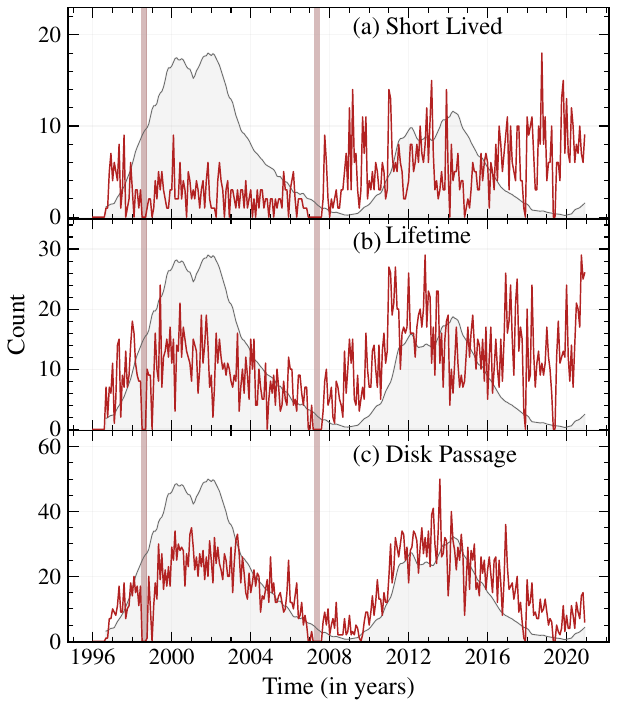}
\caption{Time series of the monthly number of unique BMRs for (a) SL, (b) LT, and (c) DP classes. The grey-shaded region represent the international smoothed sunspot number (scaled to BMR counts for better representation) \anu{from WDC-SILSO, Royal Observatory of Belgium, Brussels}. Here, vertical strips represent the data gaps in MDI magnetograms.}
\label{fig:fig6}
\end{figure}

The first property that we looked for is, does the number of newly emerging BMRs obeys the well-known solar cycle behavior. In contrast to the past studies \citep[e.g.,][]{Stenflo2012, Jha2020}, where the same BMR is counted multiple times during their appearance on the near side, instead here, we only count each tracked BMR once. Therefore, in \autoref{fig:fig6}, we show the monthly number of newly emerging BMRs with time for all three classes, SL (\autoref{fig:fig6}a), LT (\autoref{fig:fig6}b) and DP (\autoref{fig:fig6}c). The number of BMRs in LT and DP classes obediently follows the known solar cycle behavior based on the conventional sunspot number. Such behavior is not evident in the case of the SL class. A significant number of SL BMRs are also observed during the solar minima when the toroidal field of the Sun is weak and only a few large BMRs are produced.

\subsubsection{Latitude-time Distribution}
\begin{figure}[!htbp]
\centering
\includegraphics[width=0.48\textwidth]{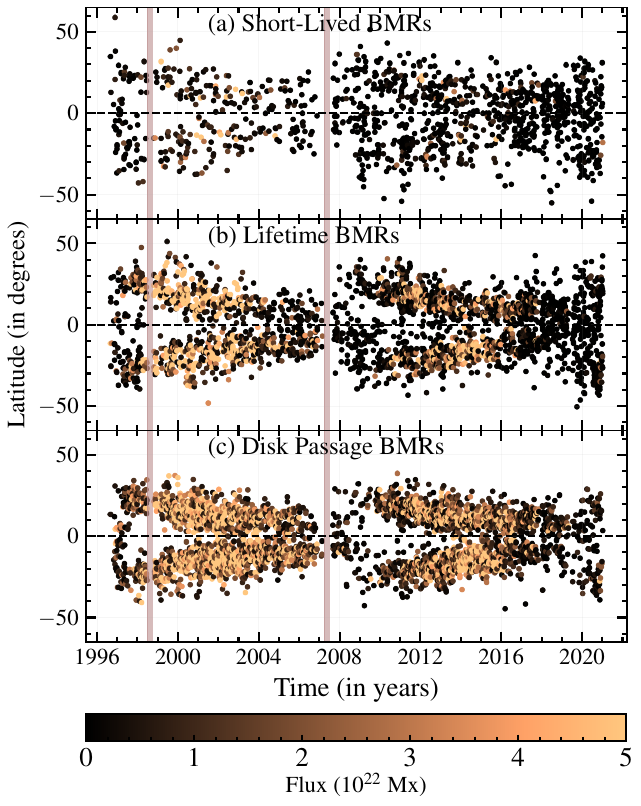}
\caption{Latitude-time distribution (butterfly diagram) of the tracked BMRs of (a) SL (b) LT, and (c) DP classes. The color of the points represents the total flux of the BMR. The vertical strips represent the data gaps in MDI magnetograms.
%\anu{and black dashed line  marks the equator}.
}
\label{fig:fig7}
\end{figure}

Another crucial property of sunspots is their latitudinal distribution \citep{Hathway2015, Jha2022}, the so-called `butterfly diagram'. In \autoref{fig:fig7}, we show the time-latitude distribution of identified BMRs for each classes.
%, SL (\autoref{fig:fig7}a), LT (\autoref{fig:fig7}b) and DP (\autoref{fig:fig7}c). 
Here, each point represents a unique BMR. The latitude and time of all the BMRs are chosen at a time when they attain their maximum total flux (represented by the colors of the points) during the tracking period. We note that for the DP class, the maximum flux may not be the actual maximum flux of the BMR, as AutoTAB only tracked them during a fraction of their life span. The interesting thing to note here is that although the BMR classes LT (\autoref{fig:fig7}b) and DP (\autoref{fig:fig7}c) follow the well-known butterfly diagram, the SL (\autoref{fig:fig7}a) class is more scattered over the latitude (a small latitudinal dependency can still be seen) independent of the phase of the solar cycle. It could be that SL class BMRs that appears  from the high latitude at the beginning of the cycle and continues till the end of the cycle at the solar minimum are part of the extended solar cycle \citep{McIntosh15} 
\bbk{and they are formed through shredding and tangling of the large-scale magnetic field itself \citep{KB16}.}
However, \citet{Jha2020} have shown that the two classes of BMRs, namely the BMRs with spot and without spot in  white-light images, follow the same latitude-time distribution based on their classification. Further study is required to dig deep into the detail, which will be done in the follow-up paper.
Another point to note here is that although the total number of BMRs in Cycle 23 is less than in Cycle 24, the fraction of bigger BMRs that falls in the LT and DP classes is more in Cycle 23. 

\subsection{\anu{Comparison with BARD}}%Existing Algorithms}

%\textcolor{red}{The comparison part seems too much to me, I don't think we have to provide that much details. One paragraph will be enough in my opinion.}

\begin{figure}[!htbp]
\hspace{-0.5cm}
\includegraphics[width=0.5\textwidth]{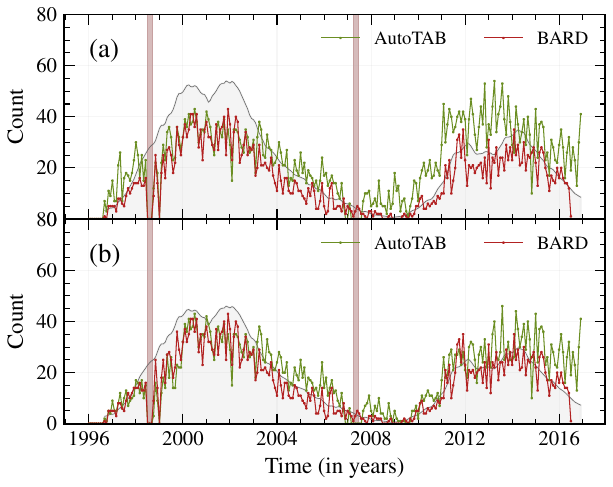}
\caption{(a) Comparison of the monthly number of newly emergent BMRs obtained from AutoTAB (our) and BARD, along with the traditional monthly sunspot number (grey shaded area enclosed by line). %\anu{from WDC-SILSO, Royal Observatory of Belgium, Brussels}.  
The sunspot number has been multiplied by a factor of 0.6 to bring it to the scale of the BMR count. The vertical strips on the figure represent the data gaps  in MDI magnetograms. \anu{(b) Same as (a) but only including the BMRs from DP class which has a lifetime of greater than one day and 500~G threshold applied to BMRs from MDI in AutoTAB dataset.}}
\label{fig:fig8}
\end{figure}

\anu{In this section, we evaluate the performance  of AutoTAB with other existing BMR tracking algorithm, BARD \citep{Jaramillodata2021, Jaramillo2021, Jaramillo2016}. To achieve this, we compute the count of newly emergent BMRs detected by both algorithms for the overlapping period, which is represented in \autoref{fig:fig8}. The number of newly emergent BMRs in BARD is obtained from BARD tilt catalog \citep[]{Jaramillo2016, Jaramillo2021, Jaramillodata2021}. 
% The catalog contains information on 9243 individually tracked BMRs, spanning the time period from 1976 to 2016.
}

\begin{deluxetable*}{rcccc}%[!t]
\tablecaption{Comparison of the statistics of new emergent BMRs between AutoTAB and BARD for different cases.
\label{tab:tab2}}
\tablehead{\colhead{ } & \multicolumn{2}{c}{Cycle 23} & \multicolumn{2}{c}{Cycle 24} \\
\colhead{ } & \colhead{AutoTAB} & \colhead{BARD}& \colhead{AutoTAB} & \colhead{BARD}}
%\colhead{} &\colhead{AutoTAB} & \colhead{BARD} & \colhead{AutoTAB} & \colhead{BARD}}
\startdata
\textbf{All BMRs} & 3861 & 2328 & 3900 & 1422\\
\textbf{BMRs with lifetime $\ge$ 1~day} & 2989 & 2328 & 2742 & 1422\\
\textbf{BMRs with lifetime $\ge$ 1~day and LT class removed (i.e.,} & 2722 & 2328 & 2114 & 1422\\
\textbf{ we are left with DP class)} &   &   &   &  \\
\textbf{All DP class BMRs plus 500~G threshold to Cycle 23} & 2649 & 2328 & 2114 & 1422\\
\enddata
\end{deluxetable*}

\anu{It can be seen from \autoref{fig:fig8}a that AutoTAB detects more BMRs throughout the intersecting period, especially during Cycle 24. As BARD has limited the cadence of the observation to one magnetogram per day, any BMR detected and tracked by BARD cannot have a lifetime less than a single day. Moreover, BARD employs different magnetic threshold values for MDI and HMI, which can bring about the disparity observed in the detected number of BMRs.  Taking these points
into consideration and 
in order to maintain the consistency between the data sets, initially, we exclude all BMRs from the SL class and any BMRs which live for less than a day from the AutoTAB data set. Nevertheless, AutoTAB still detects 28\% more new BMRs in Cycle 23 and 85\% in Cycle 24. Additionally by removing the LT class BMRs from the AutoTAB dataset and applying a 500~G magnetic threshold to the BMRs detected from MDI (Cycle 23), results in a close match of 13\% of more new BMRs detected by AutoTAB in Cycle 23. But still, we observe that in Cycle 24 AutoTAB detects 42\% of newer BMRs; see \autoref{fig:fig8}b. This suggests that BARD competently detects BMRs with bigger areas and larger flux and tends to miss smaller bipolar regions. The number of uniquely identified BMRs from both AutoTAB and BARD, over both cycles are represented in \autoref{tab:tab2}.}

\subsection{Limitations of Tracking Algorithm}
As with any algorithm, the AutoTAB also has some limitations, which are as follows.

$(i)$ One of the biggest challenges that AutoTAB has to face is dealing with the multiple BMRs lying closely together. This issue is mostly carried forward from the detection and also affects the tracking, and occasionally multiple BMRs have been tracked as a single one.

% $(ii)$ The area threshold of 150~pixels is  makes AutoTAB impossible to track smaller active regions with area $\approx150$~pixels and flux $\approx10^{20}$~Mx. These regions could be ephemeral regions with a lifetime of a few hours. These regions are ubiquitous  on the surface and may have a significant role in the understanding of BMR evolution.

$(ii)$ Once BMR crosses the west limb, the AutoTAB loses track of it, but there are a few cases, particularly for long-living BMRs, the same BMR could appear in the East limb after coming back from the far side. In this case, the AutoTAB treats this BMR as a new one and gives the unique AutoTAB-ID to it even though they are the same.   

$(iii)$ The last but not least, the detection and pre-processing steps are strongly affected by the level of noise in the data while identifying the region of interest, and ultimately it also impacts the tracking.

\subsection{Possible Application to Other Data}
 So far, we have only discussed and demonstrated the application of AutoTAB on MDI and HMI magnetogram datasets. Now  the question is, will it be possible for AutoTAB to track the BMRs in other datasets? The answer is yes, but since our algorithm has three major parts (i) detection, (ii) pre-processing, and (iii) tracking, all of them will not work in the same way with different datasets. Although the pre-processing and tracking are expected to work with various datasets, our detection algorithm, which is only optimized for MDI \citep{Stenflo2012} and HMI \citep{Jha2020} magnetograms, may not work with other data sets. Thus, for a given binary mask of the region of interest, the tracking algorithm can be used, along with pre-processing technique. Here, we would like to emphasize that even for pre-processing and tracking, a few constants or thresholds, such as morph closing kernel size, overlapping area threshold, etc., need to be tuned according to the data. \anu{For example, for the case of HMI magnetogram data with the original resolution of 4096 $\times$ 4096, the initial appropriate radius of morph closing kernel size is 16 pixels. The radius is increased to 19 pixels in three different steps as mentioned in \autoref{preprocess}. The area threshold mentioned in \autoref{preprocess} can be scaled by the resolution of the data, in the case of HMI, by a factor of 4.} Furthermore, the AutoTAB tracking algorithm can also be used to track the other features in the solar atmosphere by optimizing these parameters.

\section{Conclusions} 
\label{sec:conclusion}

Observational study of the evolving properties of BMRs is crucial to understand the origin of solar magnetic field and cycle. However, 
due to the unavailability of a completely automatic and efficient BMR tracking algorithm, the study of the temporal evolutions of various properties of BMRs is limited. 
We have developed an automatic algorithm, AutoTAB for tracking the BMR from magnetograms. It works by taking the binary maps of the detected BMR as inputs. For detecting BMRs from magnetograms, we also build an algorithm following the idea of \citet{Stenflo2012} and \citet{Jha2020}. Using our developed detection algorithm, we have produced the binary maps of the detected BMRs from HMI and MDI during the solar cycles 23 and 24. By feeding these binary maps in AutoTAB, we have produced a homogeneous and comprehensive data set for 9152 tracked BMRs.

AutoTAB is successful in tracking the identified BMRs through their life on the visible solar disk and in capturing their evolving magnetic and morphological properties. Representative examples of the dataset have been presented by showing the snapshots of the evolution of the BMRs at various stages of their life along with the progression of unsigned flux and $B_{\rm max}$.
%Further, case studies have been performed on the method's efficiency with other existing tracking methods. The study indicates that AutoTAB successfully tracks the BMRs inside the 0.9~\Rsun\ with better accuracy and consistency than the SMART. 
As AutoTAB works by checking for the overlap between consecutive binary maps, it remains independent of the dataset used and can work efficiently for tracking other solar features.
A comparison between the number of new emergent BMRs in AutoTAB and the BARD catalog shows that AutoTAB consistently detects and tracks more BMRs through the past two solar cycles compared to BARD. As AutoTAB is fully automatic, multiple closely lying BMRs might have been identified as one, which remains one of the major challenges faced by AutoTAB which needs to be addressed in the future.
%which can be accounted for the lesser number of BMRs in the maximum phase. Meanwhile, BARD uses human intervention to overcome this problem. This remains one of the major challenges faced by AutoTAB which needs to be addressed in the future.

AutoTAB tracks the BMR in a wide range of its lifetime, from hours to days, which leaves us with a large volume of tracked information. Hence, the tracked BMRs have been classified into different groups, namely, Short Lived (lives for less than 8 hours), Lifetime (emerges and disperses in the visible surface), and Disk Passage (coming from/going to the farside of the Sun). The tracked BMRs show the usual signatures of solar cycle variation. We also observe that the Lifetime and Disk Passage BMRs follow familiar latitudinal and temporal distributions, as seen by the sunspot butterfly diagram. This distribution is not observed in the case of Short Lived BMRs as they are the small features that appear at all times on the Sun. In a follow-up publication, we shall further exploit the results of tracked BMRs to compute various properties of BMR, which will help us to identify the origin of BMR formation.  \anu{The tracked information of the BMRs used in this study, along with the code of the algorithm, will be publicly accessible at https://github.com/sreedevi-anu/AutoTAB, soon after our follow-up article.}

\begin{acknowledgments}
\anu{Authors would like to thank anonymous referee for critical comments which helped in improving the manuscript.
Authors also thank Lisa Upton for reviewing the manuscript.}
AS sincerely express her gratitude to ARIES, Nainital for the warm hospitality extended during the project's initial phase.
BBK acknowledges the financial support from the Department of Science and Technology (SERB/DST), India, through the Ramanujan Fellowship (project no SB/S2/RJN-017/2018) and ISRO/RESPOND (project no ISRO/RES/2/430/19-20). 
The observational data, the LOS magnetograms from MDI and HMI, used in this article is obtained via JSOC, by the courtesy of NASA/SOHO and NASA/SDO science team. 
The BARD catalog of BMRs was downloaded from the solar dynamo dataverse (https://dataverse.harvard.edu/dataverse/solardynamo), maintained by Andrés Muñoz-Jaramillo. Sunspot data has been taken from the World Data Center SILSO, Royal Observatory of Belgium, Brussels.
\end{acknowledgments}
% \clearpage

%see \Sec{sec:app} and \Fig{}.
\appendix
\section{Comparison of the tracking results from MDI and HMI magnetograms} 
\label{sec:app}
\begin{figure*}[!htbp]
\gridline{\fig{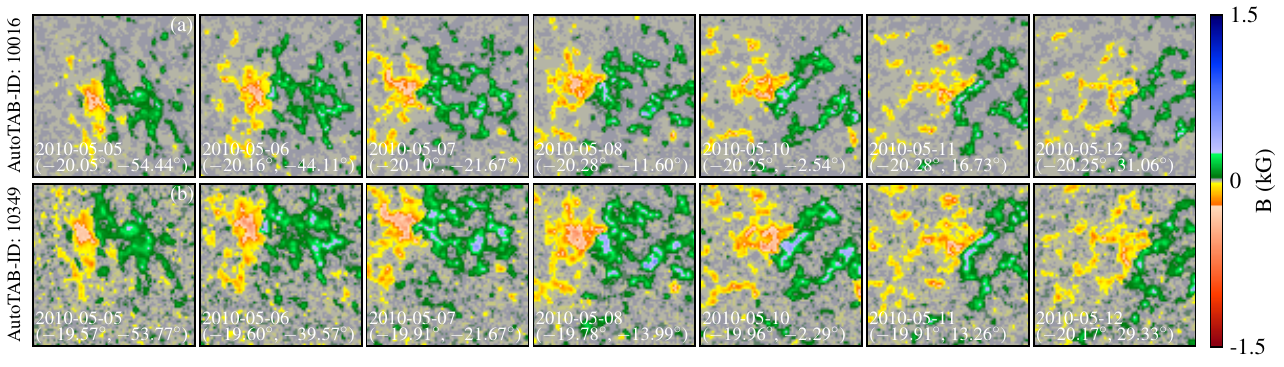}{\textwidth}{}}
\gridline{\fig{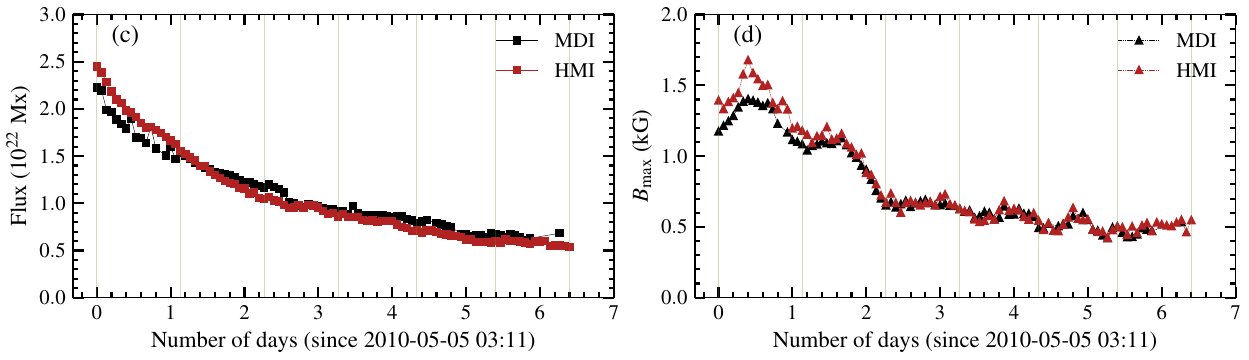}{\textwidth}{}}
     \caption{Snapshot of a tracked BMR with NOAA AR11068 during its passage on the observable disk as observed in (a) HMI and (b) MDI.
     Panels (c) and (d) show the variations of the total flux and \bmax, respectively, obtained from  both the data sets. Vertical lines in (c) and (d) represent the times corresponding to the snapshots shown in (a) and (b),
     respectively.}
     \label{fig:fig9}
\end{figure*}

Panels (a) and (b) in \Fig{fig:fig9} show the evolution of NOAA AR11068 from HMI and MDI, respectively. AutoTAB tracked this particular BMR  
for a total duration of 153 hours from 05-05-2010 19:10 to 12-05-2010 04:46, during its passage across the solar disk, with AutoTAB-ID 10349. 
As this BMR has been identified in the period when we have observations from both the instruments (HMI and MDI), we track it in both data sets for comparison. \anu{This BMR was tracked for an extra of 6 hours and more consistently by HMI (99 detections) than MDI with only 87 detections.}
%Though this BMR was tracked for the same period of time in both datasets, AutoTAB could track it more consistently with HMI data (97 detections) compared to MDI, with only 87 detections.
Less detection in MDI is expected because of the relatively larger noise in the data as compared to HMI data.  
Nevertheless,
upon closer examination of the evolution of total flux and \bmax\ throughout the lifespan of BMR (as shown in \autoref{fig:fig9}c-d), we observe that the \bmax\ and absolute total flux show excellent agreement in both the data sets. For this particular BMR, the total flux consistently decreases over the lifetime of the BMR, while \bmax\ show a small increase in the first few hours and subsequently decreases. This small increase could be due to the uncertainty in the measurement of \bmax\ near the limb.

%\section{EXAMPLES FOR DETECTION ALGORITHM}
%\anu{Detected BMRs in the month of May for the years 2010-2016 were manually checked for faulty detections. Few representative examples of detection, during the mentioned time perios is represented in \autoref{fig:app_fig2}. Here, the newly emergent BMRs for the time are shown inside black boxes, and we notice that the regions identified are indeed small BMRs that hold the applied flux balance condition.}

%\begin{figure*}[!htbp]
%    \includegraphics[width=\textwidth]{figures/ap_fig2.pdf}
%     \caption{Examples of the newly detected small BMRs (represented in black boxes) over the magnetogram for the month of May 2014 }
%     \label{fig:app_fig2}
%\end{figure*}

\clearpage
\bibliographystyle{aasjournal}
\bibliography{bmr_tracking}{}

\end{document}